\documentclass{ifacconf}

\usepackage{graphicx}      
\usepackage{natbib}        
\usepackage{amsmath,amssymb}
\usepackage{xcolor}
\usepackage{algorithm}
\usepackage{algpseudocode}
\usepackage{multicol}
\usepackage{lipsum}
\usepackage{float}

\def\QED{~\rule[-1pt]{5pt}{5pt}\par\medskip}

\newcommand{\tm}{\mathcal A}
\newcommand{\D}{\mathcal D}
\newcommand{\sm}{\bar{\mathcal A}}
\newcommand{\V}{\mathtt{V}}

\renewcommand{\Re}{\mathbb{R}}
\newcommand{\N}{\mathcal{N}}

\newcommand{\cbt}{\mathcal C_b(\mathbb{T})}
\newcommand{\lbt}{\mathcal L_\infty(\mathbb{T})}
\DeclareMathOperator*{\argmin}{arg\,min}
\DeclareMathOperator*{\argmax}{arg\,max}
\newtheorem{Assumption}{Assumption}
\begin{document}
\begin{frontmatter}

\title{Parameter Identification for Electrochemical Models of Lithium-Ion Batteries Using Bayesian Optimization} 


\author[ece]{Jianzong Pi} 
\author[me]{Samuel Filgueira da Silva} 
\author[me]{Mehmet Fatih Ozkan}
\author[ece]{Abhishek Gupta}
\author[me]{Marcello Canova}

\address[ece]{Department of Electrical and Computer Engineering, The Ohio State University, Columbus, OH, (e-mail: pi.35@osu.edu, gupta.706@osu.edu)}
\address[me]{Department of Mechanical and Aerospace Engineering, The Ohio State University, Columbus, OH, (e-mail: filgueiradasilva.1@osu.edu, ozkan.25@osu.edu, canova.1@osu.edu)}

\begin{abstract}                
Efficient parameter identification of electrochemical models is crucial for accurate monitoring and control of lithium-ion cells. This process becomes challenging when applied to complex models that rely on a considerable number of interdependent parameters that affect the output response. Gradient-based and metaheuristic optimization techniques, although previously employed for this task, are limited by their lack of robustness, high computational costs, and susceptibility to local minima. In this study, Bayesian Optimization is used for tuning the dynamic parameters of an electrochemical equivalent circuit battery model (E-ECM) for a nickel-manganese-cobalt
(NMC)-graphite cell. The performance of the Bayesian Optimization is compared with baseline methods based on gradient-based and metaheuristic approaches. The robustness of the parameter optimization method is tested by performing verification using an experimental drive cycle. The results indicate that Bayesian Optimization outperforms Gradient Descent and PSO optimization techniques, achieving reductions on average testing loss by \textcolor{black}{28.8\%} and \textcolor{black}{5.8\%}, respectively. Moreover, Bayesian optimization significantly reduces the variance in testing loss by 95.8\% and 72.7\%, respectively.


\end{abstract}


\end{frontmatter}

\section{Introduction}

Lithium ion (li-ion) batteries are at the forefront of energy storage technology and play an important role in promoting electrification in transportation (\cite{cano2018batteries,CHEN20194363}). These energy storage devices have had a significant impact on industries such as consumer electronics and electric vehicles because of their high energy density, long operating lifespan, and low maintenance requirements (\cite{xu2023high,buhmann2023consumers}). An in-depth understanding and accurate modeling of li-ion battery behavior is essential for optimizing performance and ensuring reliability in practical applications. 

Physics-based (electrochemical) battery models are essential tools for understanding, accurately predicting and monitoring lithium-ion batteries (\cite{wu2021evaluation}), as they employ sophisticated mathematical formulations to elucidate the physical mechanisms governing mass and charge transport within the li-ion cell electrodes and electrolyte solution. Parameters in electrochemical models are linked to physical, chemical and material properties of the cell components, such as electrolyte diffusion coefficient and rate constants, and their numerical values are crucial as they directly influence the accuracy and ability of the model to predict the cell behavior (\cite{seals2022physics}). By fine-tuning these parameters, electrochemical models can more precisely replicate the physical and chemical processes within the cell, ultimately leading to improved prediction of voltage and capacity, enhancing the overall battery performance and reliability.



On the other hand, parameter identification in electrochemical models presents significant challenges, notably the large number of parameters that need to be identified simultaneously, as exemplified in studies like (\cite{FORMAN2012263}).  Traditional numerical gradient-based optimization algorithms (\cite{boyd2004convex}) are commonly used in data-driven optimization tasks, but they exhibit two main drawbacks. First, they are prone to converging to local minima, with their effectiveness significantly dependent on the initial point in the descent - this compromises the robustness of the tuning process subject to the initialization. Secondly, calculating the gradient (or sub-gradient) numerically can be computationally intensive and time consuming when the objective function is complex to evaluate. More recently, reinforcement learning algorithms have shown promise in identifying optimal parameters in real-time through interaction with the system (\cite{unagar2021learning}). However, a significant drawback of these algorithms is their reliance on large datasets, which, in the context of our specific application, are impractical to obtain within a reasonable timeframe. This highlights the need for more efficient and adaptable optimization strategies in this domain.

Numerous model-free optimization techniques, including evolutionary algorithms (\cite{li2016novel}), and particle swarm optimization (PSO) (\cite{varga2013improvement, noel2012new, miranda2023particle, cavalca2018gradient, dangwal2021parameter}), were employed to improve both the speed of convergence and the resilience of model calibration. The primary hurdles include the considerable computational time required for function evaluation and gradient computation, the chance of converging to local minima, and the robustness of convergence points to randomly initialized optimization points.

Bayesian optimization (\cite{mockus2005bayesian}) is a powerful tool for globally optimizing black-box functions that come with the challenge of being computationally expensive to evaluate. Bayesian optimization provides an efficient and systematic approach to finding the global optimum by leveraging probabilistic models to systematically explore and exploit within the searching domain. Bayesian optimization enables the optimization of complex systems efficiently and robustly. The method has been widely applied to hyperparameter tuning in \cite{wu2019hyperparameter, cho2020basic}, model selection in \cite{malkomes2016bayesian}, robotics control in \cite{martinez2017bayesian}, and so on. In these applications, the key advantage of Bayesian optimization is its ability to provide robust and high-quality solutions with fewer evaluations of the objective function, which is particularly beneficial in situations where each evaluation is costly and time consuming.

The goal of this study is to apply the Bayesian optimization approach to the multi-parameter optimization problem for electrochemical models of li-ion cells, and test its performance and robustness. The study utilizes a reduced-order electrochemical model adapted from \cite{seals2022physics}. The analysis is conducted considering also gradient-based approaches (such as gradient descent) and model-free methods (such as PSO), evaluating their relative performance and robustness in the parameter identification tasks.

\section{Overview of li-ion Cell Model Equations}

In this work, the Electrochemical Equivalent Circuit Model (E-ECM) is used as a computational-friendly model to reduce computation time when performing parameter identification, resulting in shorter optimization process (\cite{seals2022physics}). The E-ECM is based on order reduction, linearization and simplification of the governing equations of the Extended Single-Particle Model (ESPM), leading to a mathematical form that allows for fast computing while maintaining the relationship between the model parameters and physical processes described by the original model equations. The constitutive equations of the E-ECM are summarized for clarity. Readers can refer to \cite{seals2022physics} for a complete description of the assumptions and derivation framework. 

The terminal voltage $V$ of the E-ECM model is given by:

\begin{equation} \label{eq:1}
\begin{split}
V(t) = U_p(c_{se,p},t) - U_n(c_{se,n},t) - (\eta_p(c_{se,p},t) \\ - \eta_n(c_{se,n},t) )+ \phi_{diff}(t) + \phi_{ohm}(t) - I(t)R_c
 \end{split}
\end{equation}




The kinetic overpotential $\eta_i$ and the exchange current density $i_{0i}$ are defined as:

\textcolor{black}{\begin{equation} \label{eq:6}
\begin{split}
\eta_{i}(t) = \frac{\bar R T_0(-J_iI(t))}{Fi_{0,i}}
 \end{split}
\end{equation}}

\textcolor{black}{\begin{equation} \label{eq:5}
\begin{split}
i_{0i}(t) = exp\left( \left(  \frac{1}{T_{ref}} - \frac{1}{T(t)} \right) \frac{E_{ioi} }{\bar R}   \right) F k_i \sqrt{c_i(c_{max,i} - c_i)c_{e,i}}
 \end{split}
\end{equation}}

where $i=p,n$. The Pade approximation is used to calculate the lithium surface concentration $c_i$ for the solid phase diffusion:

\textcolor{black}{\begin{equation} \label{eq:7}
\begin{split}
c_i(s) = c_{i0} + (G_b(s) + G_d(s))\frac{-R_i}{3F\epsilon_{am,i}L_iA} I(s)
 \end{split}
\end{equation}}

where the transfer function for bulk concentration $G_b(s)$ and the transfer function for diffusion dynamics $G_d(s)$ are expressed as:

\textcolor{black}{\begin{equation} \label{eq:8}
\begin{split}
G_b(s) = \frac{\frac{2}{7}\frac{R_i}{D_i}s + \frac{3}{R_i}}{\frac{1}{35}\frac{R_i^2}{D_i}s^2 + s}
 \end{split}
\end{equation}}

\textcolor{black}{\begin{equation} \label{eq:9}
\begin{split}
G_d(s) = \frac{\frac{R_i}{5D_i}}{\frac{1}{35}\frac{R_i^2}{D_i}s + 1}
 \end{split}
\end{equation}}

The potential drop in the electrolyte solution, $\phi_e$, is analogously given by:

\textcolor{black}{\begin{equation} \label{eq:10}
\begin{split}
\phi_e(s) = (G_{pos}(s) + G_{neg}(s))\frac{C_1}{D_{e}}I(s)
 \end{split}
\end{equation}}

\begin{figure*}[h!]
    \begin{center}
       \includegraphics[angle=0,scale=0.55]{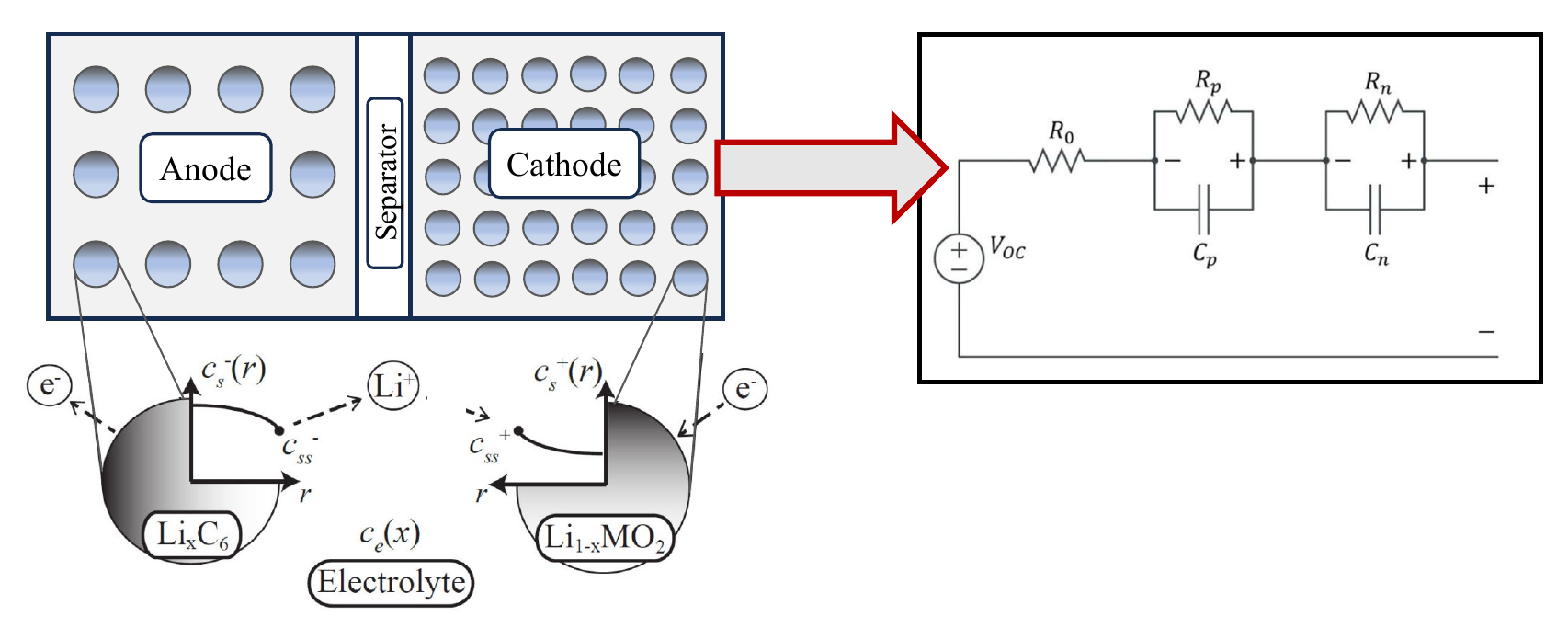}
    \caption{Schematic of the E-ECM model.}
    \label{fig:1}
    \end{center}
\end{figure*}

where $C_1$ is defined as a function of the cell parameters, as shown in Eq. \ref{eq:13}, and $G_{pos}$ (Eq. \ref{eq:11}) and $G_{neg}$ (Eq. \ref{eq:12}) are positive and negative transfer functions.

\begin{figure*} [!h]
\par\noindent\rule{\dimexpr(0.5\textwidth-0.5\columnsep-0.4pt)}{0.4pt}%
\rule{0.4pt}{6pt}
\begin{equation} \label{eq:13}
\textcolor{black}{C_1 = 2RT\left(\frac{-L_{cell}}{A_s}\frac{1}{F^2c_{e0}}\right)(1-t_0^{+})(1+\beta)\left(0.601 -
0.24 \sqrt{\frac{c_{e0}}{1000}} +  0.982\left(1 - 0.0052(T_0-T_{ref}) \left(\frac{c_{e0}}{1000} \right)^{1.5}  \right)\right)}
\end{equation}
\vspace{\belowdisplayskip}\hfill\rule[-6pt]{0.4pt}{6.4pt}%
\rule{\dimexpr(0.5\textwidth-0.5\columnsep-1pt)}{0.4pt}
\end{figure*}

\textcolor{black}{\begin{equation} \label{eq:11}
\begin{split}
G_{pos}(s) = \frac{0.124\gamma_p}{\frac{0.1052L_{cell}^2}{D_{e}} s + 1}
 \end{split}
\end{equation}}

\textcolor{black}{\begin{equation} \label{eq:12}
\begin{split}
G_{neg}(s) = \frac{0.117\gamma_n}{\frac{0.0997L_{cell}^2}{D_{e}} s + 1}
 \end{split}
\end{equation}}









Last, the potential drop due to ohmic losses, $\phi_{ohm}$, is expressed by Eq. \ref{eq:14}, where $\kappa$ is a function of the initial electrolyte concentration $c_{e0}$ and cell temperature $T$.

\textcolor{black}{\begin{equation} \label{eq:14}
\begin{split}
\phi_{ohm}(t) = \frac{-I(t)L_{cell}}{\kappa A}
 \end{split}
\end{equation}}

Sensitivity analysis has been previously used to rank the parameters of li-ion battery models with respect to their effect on the terminal voltage output response (\cite{jin2018parameter,park2018optimal}). In \cite{dangwal2021parameter}, seven parameters are highlighted for holding high influence on model response in transient conditions. In this current work, three of those parameters are evaluated for the implementation of the multi-parameter identification process: electrolyte diffusion coefficient $D_e$, cathode rate constant $k_p$, and anode rate constant $k_n$.



\section{Bayesian Optimization}
Bayesian Optimization is an efficient method for finding the maximum or minimum of an objective function that is expensive to evaluate. The method combines a probabilistic model, typically a Gaussian Process, to estimate the function and an acquisition function to decide where to sample next. This approach effectively balances exploration of unknown areas and exploitation of promising regions, making it ideal for tasks like hyperparameter tuning in machine learning, where evaluations (such as simulating a high-fidelity model) are costly and time-consuming. 

\subsection{Notations}
Let $\Theta\subset\Re^n$ be the parameter space and 
$\mathbb T:= [0, T]$.
Let $\lbt$, $\cbt$ be the collection of almost everywhere bounded and continuous bounded functions defined on $\mathbb T$ respectively. 

\subsection{Preliminaries}
\begin{fact}[Conditional Multivariate Gaussian] Consider two multivariate Gaussian random variables $X \sim \N(\mu_X, \Sigma_X) \in \mathbb R^p$ and $Y\sim \N(\mu_Y, \Sigma_Y) \in \mathbb R^m$. Then the distribution of $X|(Y=y)$ is multivariate Gaussian with mean 
    \begin{align*}
        \text{mean vector = } &\mu_X + \Sigma_{XY}\Sigma_{YY}^{-1}(y - \mu_Y), \\
        \text{covariance matrix = } &\Sigma_{XX} - \Sigma_{XY}\Sigma_{YY}^{-1}\Sigma_{YX},
    \end{align*}
where $\Sigma_{XY} \in \mathbb R^{p\times m}$, $\Sigma_{YX} \in \mathbb R^{m\times p}$ are the covariance matrices between $X$ and $Y$.
\label{fact:conditionalGaussian}
\end{fact}

\subsubsection{Gaussian Process}
Let $f: \Theta \rightarrow \mathbb R$ be the objective function. A Gaussian process over domain $\Theta$ is a collection of random variables $(f(\theta))_{\theta \in \Theta}$ such that each finite subcollection $(f(\theta_i))_{i=1}^N$ are jointly Gaussian with mean $E[f(\theta_i)] := \mu(\theta_i)$ and covariance $E[(f(\theta_i) - \mu(\theta_i)(f(\theta_j)-\mu(\theta_j))] := k(\theta_i, \theta_j)$. Let $k$ be the \textit{kernel} function that determines the covariance structure. Assume $k:\Theta \times \Theta \rightarrow \mathbb R$ to be positive definite: the kernel evaluation matrix $[k(\theta_i, \theta_j)]_{i, j} \in \mathbb R^{N\times N}$ is positive semi-definite for every finite collection $(\theta_i)_{i=1}^N$. In practice, one may pick $k$ to be the Gaussian kernel (or the exponential squared function):
$$k(\theta, \theta') = \exp\left(-\frac{\|\theta - \theta'\|^2}{2}\right).$$
The Gaussian process defined above is denoted as $$GP(\mu(\cdot), k(\cdot, \cdot)).$$ The process of inference with function evaluations is as follows:
first, set the prior of $f$ to be $GP(0, k(\cdot, \cdot))$. Then, randomly pick $s$ points $(\theta_1, \dots \theta_s)$, run the function evaluations on $\theta_1, \dots \theta_s$ and record $f(\theta_1), \dots, f(\theta_s)$. Then, one can construct the kernel covariance matrix $K \in \mathbb R^{s\times s}$, where $[K]_{ij} = k(\theta_i, \theta_j)$. 

For a new point $\theta_{s+1}$, Fact \ref{fact:conditionalGaussian} implies $f(\theta_{s+1})$ is Gaussian with mean $\mu(\theta_{s+1}) = \mathbf k^TK^{-1}[f(\theta_1), \dots, f(\theta_s)]^T$, and variance $\sigma^2(\theta_{s+1}) = k(\theta_{s+1}, \theta_{s+1}) -\mathbf k^TK^{-1}\mathbf k$. Where the vector $\mathbf k := [k(\theta_1, \theta_{s+1}), \dots, k(\theta_s, \theta_{s+1})]^T \in \mathbb R^s$. See \cite[pp.16 - 26]{garnett2023bayesian} for a detailed discussion of inference with Gaussian processes.

\subsubsection{Aquisition Function}
Let $\D$ be the \textit{dataset} containing observations $\{\theta_i, f(\theta_i)\}_i$ so far. Let $a:\Theta \to\mathbf R^+$ be the \textit{acquisition function} that determines which parameter to be evaluated, specifically, 
the algorithm determines the next point to evaluate by a \textit{proxy} optimization 
$\theta_{N+1} = \argmax_{\theta\in\Theta} a(\theta)$. The acquisition function that is applied in this experimental study is the expected improvement (\cite{jones1998efficient}): Suppose $\tilde f$ is the minimal value of $f$ observed so far, then the expected improvement is defined as:
    $$a_{EI}(\theta) := E[\max(0, \tilde f - f(\theta) | \D]$$

Other commonly used acquisition functions are probability of improvement, Bayesian expected loss, Thompson sampling and so on (for further details, see \cite{frazier2018tutorial}). Combining Gaussian process and the acquisition function, the Bayesian optimization algorithm is shown in Algorithm \ref{alg:bayesopt}.

\begin{algorithm}
\caption{Bayesian Optimization}\label{alg:bayesopt}
\begin{algorithmic}
\Require Gaussian prior on $f$, acquisition function $a$.
\State Sample $s_0$ points and observe $\{\theta_i, f(\theta_i)\}_{i=1}^{s_0}$.
\For{$i = 1,\dots, S$}
    \State Update posterior on objective function $f$.
    \State Update acquisition function $a$.
    \State Choose $\theta_i = \arg\max_{\theta \in \Theta} a(\theta)$.
    \State Evaluate objective function and observe $(\theta_i, f(\theta_i))$.
\EndFor
\\
\Return{Observed $\theta_i$ that minimizes $f$.}
\end{algorithmic}
\end{algorithm}

\section{Problem Formulation}
Denote the physical Li-ion cell as $\sm: \lbt\to\cbt$ which outputs the terminal cell voltage as a time series $\V \in \cbt$ and represent an E-ECM as a nonlinear operator $\tm:\Theta\times\lbt\to\cbt$. For every fixed parameter $\theta \in \Theta$, and every input excitation $\mathbf u\in\lbt$, $\tm_{\theta}$ outputs the model terminal cell voltage of time series $\V_m \in \cbt$. Assuming that experimental input excitation profiles $\mathbf u_1, \dots, \mathbf u_K$ and $\sm(\mathbf u_1), \dots, \sm(\mathbf u_K)$ are available, the following optimization problem is formulated:
\begin{align}
    \theta^* = \argmin_{\theta \in \Theta}\sum_{k=1}^K\|\tm(\theta, \mathbf u_k) - \sm(\mathbf u_k) \|_2^2 =: f(\theta)
    \label{eqn:opt_prob}
\end{align}
Evaluation of this objective function is costly because it involves computing a time series output by solving the set of differential and algebraic equations underlying in the E-ECM. Furthermore, determining the gradients of these objective functions poses an even greater challenge, due to the large number of samples required to numerically compute the gradient with respect to the optimization variables. 

The Bayesian optimization method addresses the aforementioned challenges. Since Gaussian processes are used as the function approximator, the following assumption is required:
\begin{Assumption}
    The mapping $\theta \mapsto f(\theta) = \sum_{k=1}^K\|\tm(\theta, \mathbf u_k) - \sm(\mathbf u_k) \|_2^2$ is continuous over $\Theta$.
\end{Assumption}

\section{Case Study}
In the optimization framework defined by Equation (\ref{eqn:opt_prob}), the parameter space $\Theta$ can be either a one-dimensional interval or a multi-dimensional box. This work considers a set of three model parameters $\theta = (k_p, k_n, D_{e})$ to optimize concurrently. Thus, $\Theta$ represents a three-dimensional box containing the lower and upper limits of the parameters.

\begin{figure}[h]
    \centering
    \includegraphics[width=\linewidth]{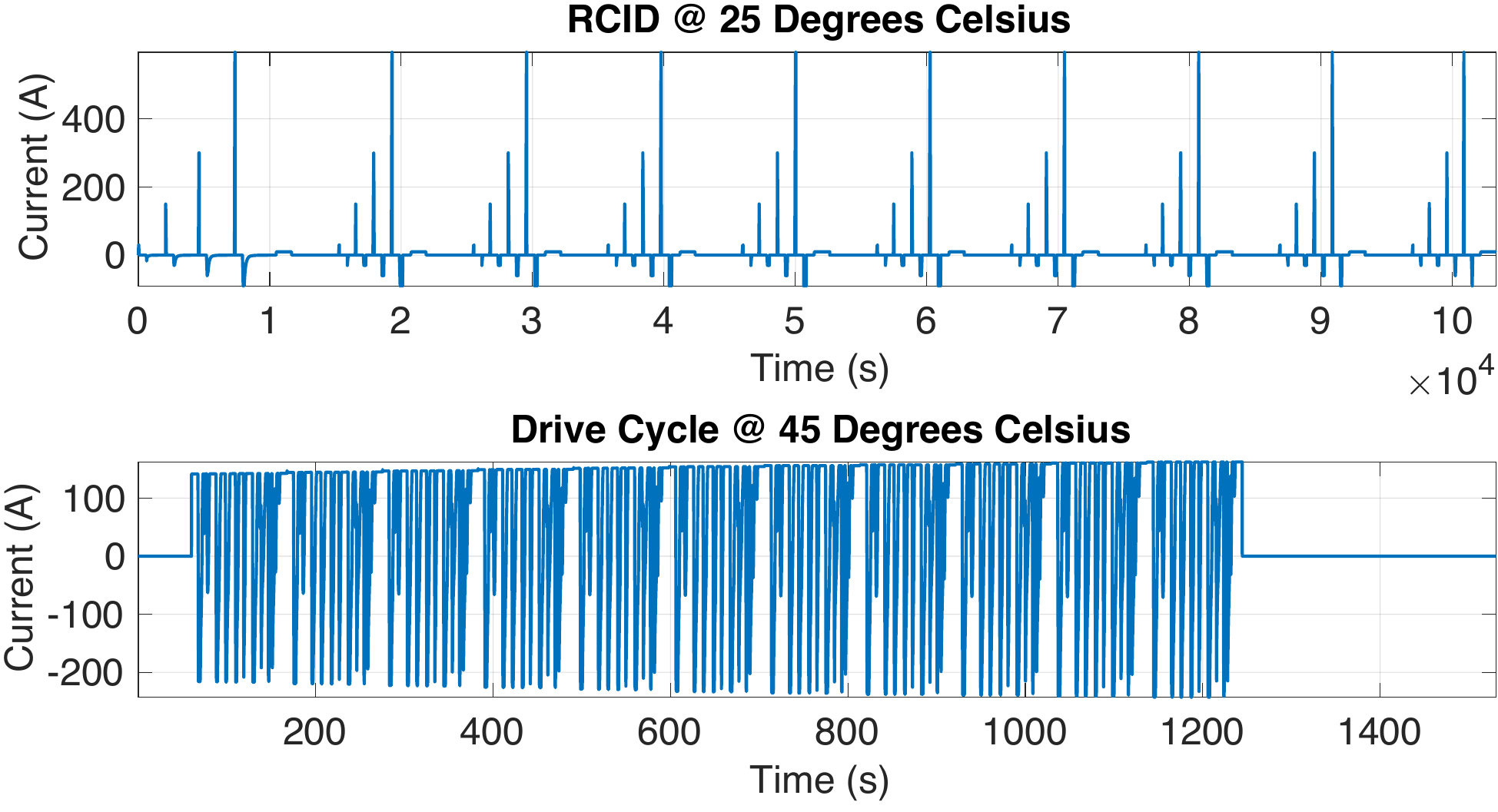}
    \caption{The RCID (training set) and drive cycle (verification set) input current profiles.}
    \label{fig:test_profiles}
\end{figure}

The optimization of the objective function in Eq. (\ref{eqn:opt_prob}) is conducted using a training set comprising a dynamic, step-wise current profile, depicted in the upper panel of Figure \ref{fig:test_profiles}. This profile, named RCID, spans the entire feasible C-rate and state of charge (SOC) range of a li-ion cell. To assess the generalization power of the optimized parameters, the results are evaluated on a different duty cycle formed by repetitions of a drive cycle profile, as shown in the lower panel of Figure \ref{fig:test_profiles}.

\section{Results and Discussion}
The training and testing losses of the Bayesian optimization method applied to the multi-parameter identification problem are evaluated by benchmarking against two common methods, namely gradient descent and PSO. For each method, the starting point is initialized uniformly at random within $\Theta$. To ensure fairness, the number of function evaluations is capped at $50$ and $10$ repetitions are executed for each method. Mean and variance values for training time, training losses, and testing losses are computed and summarized in Table 1. 
\begin{table*}[htbp] \label{table:results}
\centering
\caption{Comparison among gradient descent, PSO and Bayesian optimization in optimization time, and training and testing losses.}
\label{table:results}
\begin{tabular}{|c|cc|cc|cc|}
\hline
                      & \multicolumn{2}{c|}{Time (seconds)} & \multicolumn{2}{c|}{Training loss (V$^2$)} & \multicolumn{2}{c|}{Testing loss (V$^2$)} \\ \hline
                      & Mean          & Variance        & Mean          & Variance       & Mean          & Variance       \\ \hline
Gradient Descent      & 215.865       & 7625.6          & 33.956        & 3.662          & 0.848         & 0.071          \\ \hline
PSO                   & 580.115       & 433.45          & 32.04         & 0.233          & 0.641         & 0.011          \\ \hline
\textbf{Bayesian Optimization} & \textbf{540.395} & \textbf{17.337} & \textbf{31.556} & \textbf{0.007} & \textbf{0.604} & \textbf{0.003} \\ \hline
\end{tabular}
\end{table*}

\begin{figure}
    \centering
    \includegraphics[width=1\linewidth]{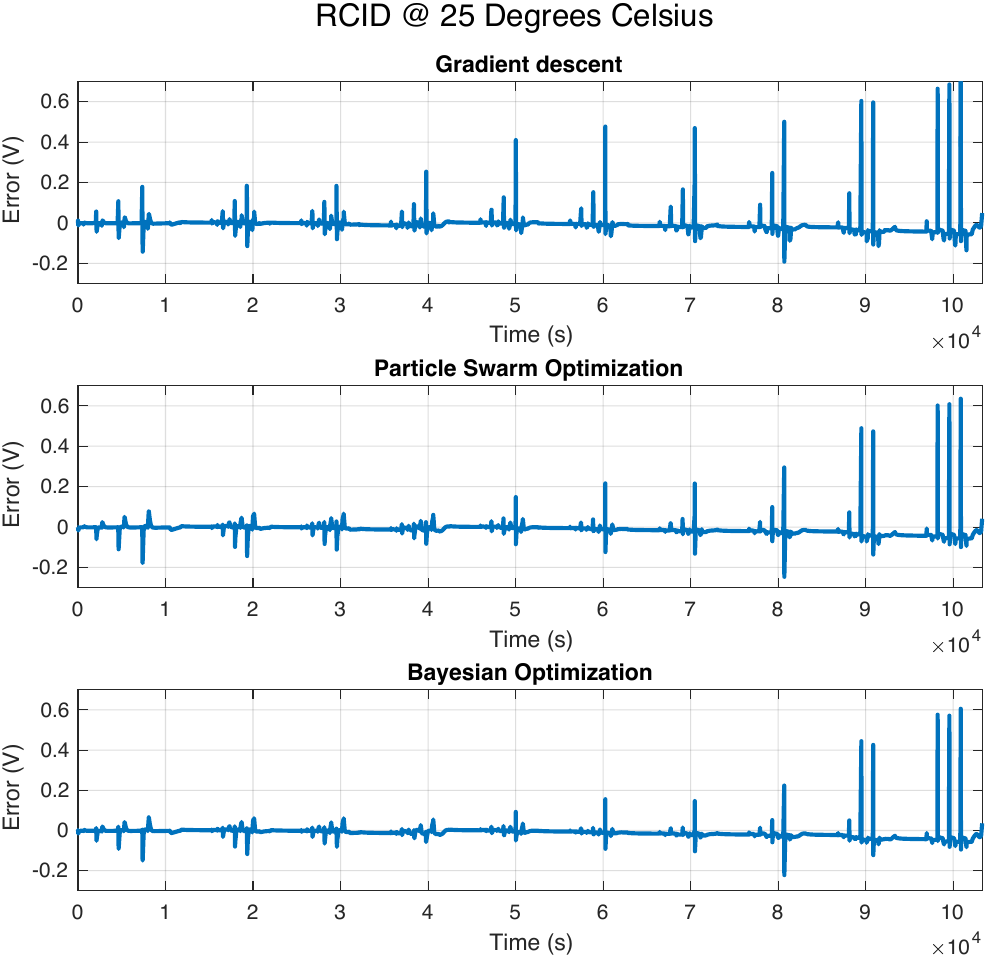}
    \caption{Cell terminal voltage error in RCID test among converged parameters in gradient descent, PSO and Bayesian optimization (data collected at $T=25^\circ C$).}
    \label{fig:rcid}
\end{figure}

\begin{figure}
    \centering
    \includegraphics[width=1\linewidth]{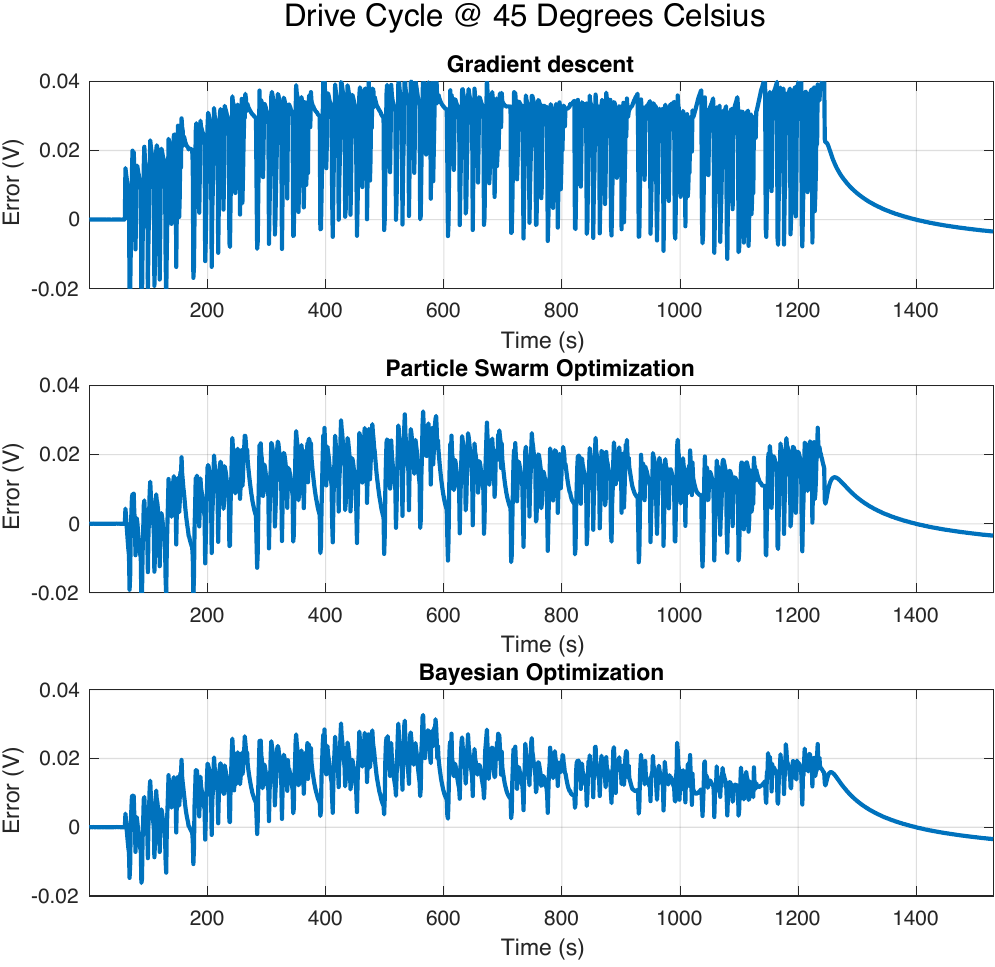}
    \caption{Cell terminal voltage error in drive cycle testamong converged parameters in gradient descent, PSO and Bayesian optimization (data collected at $T=45^\circ C$).}
    \label{fig:drive_cycle}
\end{figure}

Additionally, the voltage error for two input profiles (RCID and drive cycle) is visualized. The tuned parameters corresponding to the median training loss are utilized to generate the error plots depicted in Figures \ref{fig:rcid} and \ref{fig:drive_cycle}. It is evident that Bayesian optimization more accurately predicts the terminal voltage than gradient descent and PSO methods, due to its superior tuning of the electrochemical parameters in these tests.

\begin{figure}[h!]
    \begin{center}
       \includegraphics[angle=0,scale=0.55]{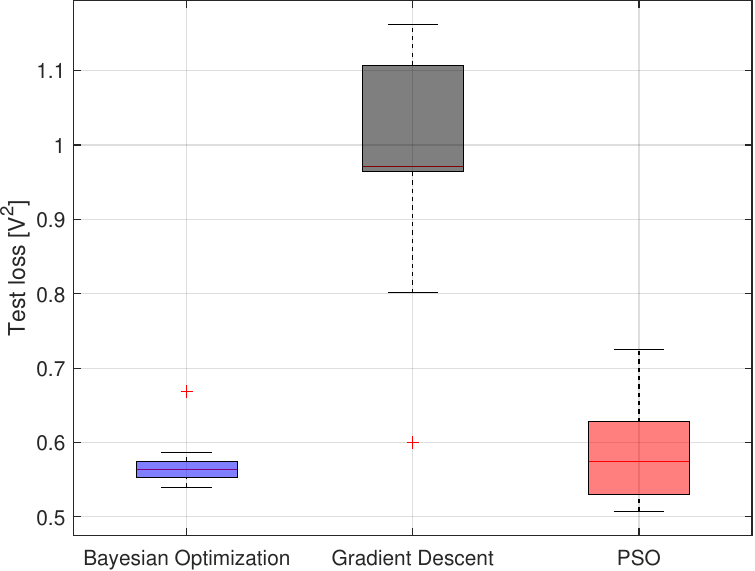}
    \caption{Box plot for the testing losses.}
    \label{fig:boxplot}
    \end{center}
\end{figure}

As depicted in the Table 1, the Bayesian optimization method surpasses both gradient descent and PSO in both training and testing losses, with improvements of 28.8\% and 5.8\% on average, respectively. Moreover, Bayesian optimization significantly reduces the variance in the training and testing losses by 95.8\% and 72.7\%, respectively. 

To further showcase the robustness of the Bayesian optimization method on test sets, the box plot in Figure \ref{fig:boxplot} illustrates the testing loss evaluated across the 10 repetitions for the drive cycle. The results indicate that the gradient descent method shows higher mean, median losses, and greater variability compared to the Bayesian optimization and the PSO methods, indicating less consistent performance across test cases. On the other hand, both the Bayesian optimization and PSO have narrower ranges compared to gradient descent, indicating more stable performance within a tighter range of losses. It is worth highlighting that the Bayesian optimization method demonstrates better performance consistency compared to PSO by showing less variability in testing losses. 

The findings collectively demonstrate that Bayesian optimization emerges as a promising approach for multi-parameter identification of the electrochemical battery model, excelling in both performance and robustness.


\section{Conclusion}
This study addresses the challenge of multi-dimensional parameter tuning in electrochemical battery modeling by formulating the problem as a large-scale optimization and applying the Bayesian optimization method.

 Compared to conventional methods such as particle swarm optimization (PSO) and gradient descent, Bayesian optimization exhibits significant improvements. It not only enhances training performance and robustness but also reduces training losses more effectively—achieving average improvements of 28.8\% over gradient descent and 5.8\% over PSO. Furthermore, Bayesian optimization demonstrates a reduction in variance by 95.8\% and 72.7\% relative to gradient descent and PSO, respectively. These findings underscore the potential of Bayesian optimization as a robust algorithm for parameter tuning in battery models.

Potential future work may include expanding the optimization to larger sets of parameters. Furthermore, the Bayesian optimization algorithm could be augmented by integrating sensitivity information in the objective function. As indicated by \cite{makrygiorgos2023gradient, ament2024unexpected}, leveraging sensitivity information can enhance Bayesian optimization by adjusting acquisition functions and altering the approach to exploring new points. By incorporating these concepts, there is potential to accelerate the optimization process, ultimately improving improve both performance and robustness in the multi-parameter optimization task.

\begin{ack}
The authors wish to acknowledge the Honda Research Institute for the inspiring discussions and helpful feedback that contributed to the research presented in this paper.
\end{ack}

\bibliography{ifacconf}       
\end{document}